\documentclass{elsart1p}
\usepackage{epsfig}
\usepackage{ulem,color,graphics}
\usepackage{amssymb}
\usepackage{graphicx}
\usepackage{epsfig}
\usepackage{bm}

\def\etal{{\it et al.}}
\def\go{\rightarrow  }
\def\be{\begin{equation}}
\def\ee{\end{equation}}
\def\br{\begin{eqnarray}}
\def\er{\end{eqnarray}}
\def\brn{\begin{eqnarray*}}
\def\ern{\end{eqnarray*}}
\def\rf#1{{(\ref{#1})}}
\newcommand{\tLN}{\mathsf{t}_{\Lambda N}}
\def\I {{{\cal I}}}
\def\M {{{\cal M}}}
\def\O {{{\cal O}}}

\def\T {{{\cal T}}}
\newcommand{\TI}{\mathsf{T}}

\def\bit{\begin{itemize}}
\def\eit{\end{itemize}}
\def\Ket#1{||#1 \rangle}
\def\Bra#1{\langle #1||}

\def\ie{{\it i.e., }}

\def\ket#1{|#1 \rangle}
\def\rf#1{{(\ref{#1})}}

\def\go{\rightarrow  }
\def\sqi{\frac{1}{\sqrt{2}}}
\newcommand{\Mass}{\mathrm{M}}

\def\rf#1{{(\ref{#1})}}

\def\ket#1{|#1 \rangle}
\def\Ket#1{||#1 \rangle}
\def\Bra#1{\langle #1||}

\def\be{\begin{equation}}
\def\ee{\end{equation}}
\def\br{\begin{eqnarray}}
\def\er{\end{eqnarray}}
\def\etc{ {\it etc}}

\begin{document}

\begin{frontmatter}

\title{ Nonmesonic Weak Decay Spectra and Three-Nucleon Emission}
\author[a]{Claudio  De Conti}
\author[b]{Airton Deppman}
\author[c]{Franjo Krmpoti\'c}
\address[a]{Campus Experimental de Itapeva, Universidade Estadual Paulista, 18409-010,
Itapeva, SP, Brazil}
\address[b]{Instituto de F\'isica, Universidade de S\~ao Paulo, S\~ao Paulo, Brasil}
\address[c]{Instituto de F\'isica La Plata, CONICET, 1900 La Plata,
Argentina, and Facultad de Ciencias Astron\'omicas y Geof\'isicas,
Universidad Nacional de La Plata, 1900 La Plata, Argentina.}
\thanks[cor1]{Corresponding author: Franjo Krmpoti\'c:,  e-mail address:
krmpotic@fisica.unlp.edu.ar}

\begin{abstract}
We have evaluated the nonmesonic weak decay spectra within the independent-particle shell-model, and compared them
with the recent measurements  of: i) the single and double coincidence
nucleon spectra in ${\mathrm{^{12}_{\Lambda}C}}$ performed at KEK,  and ii) proton kinetic energy spectra
in ${\mathrm{^{5}_{\Lambda}He}}$, ${\mathrm{^{7}_{\Lambda}Li}}$, ${\mathrm{^{9}_{\Lambda}Be}}$, ${\mathrm{^{11}_{\Lambda}B}}$,
${\mathrm{^{12}_{\Lambda}C}}$, ${\mathrm{^{13}_{\Lambda}C}}$, ${\mathrm{^{15}_{\Lambda}N}}$
and  ${\mathrm{^{16}_{\Lambda}O}}$ done by FINUDA.
 Based on this comparison we argue that the extraction from the data
of  the three-body $\Lambda NN \rightarrow nNN$
induced decay rate, as done in these works,  could be questionable.

\end{abstract}

\end{frontmatter}

\section{Introduction}

The weak decay rate of a $\Lambda$ hypernucleus can be cast as~\cite{Al02}
\be
\Gamma_W = \Gamma_M +\Gamma_{NM},
\label{1}\ee
where $\Gamma_M$ is  decay  rate for the mesonic (M) decay $\Lambda \rightarrow \pi N$,  and  $\Gamma_{NM}$
is the rate for the nonmesonic (NM) decay, which can be induced either by one bound nucleon ($1N$),
$\Gamma_1(\Lambda N  \rightarrow nN)$,
or by two bound nucleons ($2N$), $\Gamma_2(\Lambda NN \rightarrow nNN)$, where $N=p,n$
\ie
\be
\Gamma_{NM}=\Gamma_1+\Gamma_2; \hspace{1cm}\Gamma_{1}=\Gamma_p+\Gamma_n, \hspace{1cm}\Gamma_{2}=\Gamma_{nn}+\Gamma_{np}+\Gamma_{pp}.
\label{2}\ee

While the M and  $1N$-NM decays have been seen experimentally in the pioneering measurement
performed  more than 50 years ago by Schneps, Fry, and Swami~\cite{Sc57},
 the experimental observation of the $2N$-NM decay, which was predicted
 by Alberico \etal~\cite{Al91} in 1991
 (see also Ref.~\cite{Ra94}),
  has been reported only
quite recently at KEK~\cite{Ki09}, and  at FINUDA~\cite{Ag10}.
Both groups announced  a  branching ratio $\Gamma_{2}/\Gamma_{NM}\sim 25-30~\%$. The first group obtained this  result
from  the single and double coincidence nucleon spectra in
${\mathrm{^{12}_{\Lambda}C}}$, and the second one from proton
kinetic energy spectra in ${\mathrm{^{5}_{\Lambda}He}}$,
${\mathrm{^{7}_{\Lambda}Li}}$, ${\mathrm{^{9}_{\Lambda}Be}}$,
${\mathrm{^{11}_{\Lambda}B}}$, ${\mathrm{^{12}_{\Lambda}C}}$,
${\mathrm{^{13}_{\Lambda}C}}$, ${\mathrm{^{15}_{\Lambda}N}}$ and
${\mathrm{^{16}_{\Lambda}O}}$.
A branching ratio for the $2N$-NM decaying channel
of such a magnitude is consistent with the prediction made by  Bauer and Garbarino~\cite{Bau09}, but it is
very large in comparison with
the upper limit $\Gamma_{2}/\Gamma_{W}\le 0.097$ ($95\%$  CL),
established previously from
the single-particle proton and neutron kinetic energy spectra
in ${\mathrm{^{4}_{\Lambda}He}}$ at BNL~\cite{Pa07}.
In this  experiment it was also found that the effect of
the final state interactions (FSIs) is relatively small,  its upper limit being
$\Gamma_{NM}^{FSI}/\Gamma_{W}\le 0.11$ ($95\%$  CL).
\footnote{In fact,  the ${\mathrm{^{4}_{\Lambda}He}}$ spectra are accounted for reasonably well
theoretically by considering the $1N$-NM decay mode only~\cite{Ba09}.}

Quite recently, Bauer and Garbarino~\cite{Bau11} have obtained good   agreement with   KEK data~\cite{Ki09}, considering
 both the one- and the  two-nucleon induced decays in the framework of the
 Fermi Gas Model (FGM). These authors have also analyzed the proton ${\mathrm{^{12}_{\Lambda}C}}$ spectrum
 measured at FINUDA~\cite{Ag10}, but no theoretical study of the remaining spectra  has been done so far.
The aim of the present work is twofold. First, we confront the results of
the independent particle shell model (IPSM) for the $1N$-NM decay with the above mentioned experiments.
 Second, based on this comparison we expose our doubts about the reliability
 of the procedures followed in these works to  extract
the ratio  $\Gamma_{2}/\Gamma_{NM}$  from  data.

\begin{figure}[htpb]
\begin{center}
\includegraphics[width=0.6\linewidth,clip=]{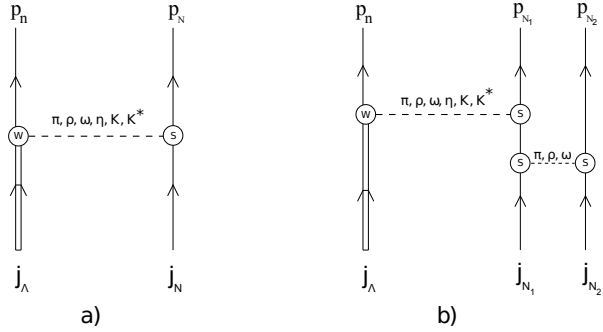}
\caption{\label{F1} Schematic representation of a)  one-nucleon,
and  b) two-nucleon induced decays in $\Lambda$-hypernuclei when
described by the interplay of weak ($W$) and  strong ($S$)
interactions through the exchanges of nonstrange-mesons
$\pi,\rho,\omega$, and $\eta$, and strange-mesons $K$, and $K^*$.}
\end{center}
\end{figure}
The  schematic representation of the two channels, when the decay dynamics is accounted for
 by the  one meson-exchange (OME),  is shown in Fig. \ref{F1}.
 This  is the most frequently used model for handling the NM-decay, and includes
  the exchanges of nonstrange-mesons $\pi,\rho,\omega$, and $\eta$, and strange-mesons $K$, and $K^*$.
It is based on the original idea of
 Yukawa that the $NN$  interaction at long distance is due to the one-pion-exchange (OPE),
the dominant role being played  by the
exchange of  pion and kaon mesons.

The OPE potential was verified quantitatively by the Nijmegen
partial wave analysis of the NN scattering in the elastic
region~\cite{St94}. \ie at distances larger than the minimal de
Broglie wavelength $1/\sqrt{m_\pi\Mass}\sim 0.5$ fm corresponding
to the pion production threshold. The verification of other meson
exchanges is less straightforward, and the uncertainties in the
baryon-baryon-meson (BBM) coupling constants  could be quite
sizeable since they are not constrained by experiments. To derive
them in the strong sector (S vertices in Fig. \ref{F1}) the
$SU(3)_f$ (flavor) symmetry is  utilized.  In the weak sector (W
vertex in Fig. \ref{F1}) the BBM parity-violating couplings are
obtained from the $SU(6)_W$ (weak) symmetry, while the parity
conserving ones are derived  from  a pole model with only baryon
pole resonances~\cite{Pa97}.

The uncertainties regarding  the BBM coupling constants are
amplified still further by the short-range correlations (SRCs) between the emitted nucleons  $nN$, and $nNN$.
 Parre\~{n}o, and  Ramos have shown that they can diminish the value of $\Gamma_1$ by more than a factor of two~\cite{Pa02}.
 Nothing has been said so far regarding the effect of the SRCs on the $2N$-NM decay.
The theoretical scene becomes still more complex when effects of quark degrees of freedom~\cite{Sa00,Sa02},
the $2\pi$-exchanges~\cite{Pa04,Ch07,It08}, and the axial-vector $a_1$-meson exchange~\cite{It08} are considered.

To derive  the ratio  $\Gamma_{2}/\Gamma_{NM}$ from  data, both  KEK~\cite{Ki09}
 and  FINUDA~\cite{Ag10} teams have
 included the FSIs in their analysis.
 The first group has used them to gauge their  experimental data on single and double coincidence nucleon spectra in
 ${\mathrm{^{12}_{\Lambda}C}}$.
For this purpose they have  used
the   Intranuclear Cascade (INC) code, which was developed by Ramos \etal~\cite{Ra97}
  to follow the fate of nucleons produced by the primary $1N$-NM and $2N$-NM
decays.
On the other hand,  the FINUDA's group~\cite{Ag10}, in their analysis of proton kinetic energy spectra
in several hypernuclei, have  assumed that the basic role of  the FSIs
is to  shift the transition strength  from high energy to low energy,
without  substantialy modifying the total number of protons.

The above mentioned experiments, together with several others performed during the last
 decade~\cite{Ki03,Ok04,Ok05,Ou05,Ka06,Ki06,Bh07,Ag08,Ag09},
represent very important  advances in  our knowledge about the NM decay.  Explicitly, they are: 1) new
high quality measurements of  single-nucleon spectra $S_{N}(E_N)$,
 as a function of one-nucleon energy $E_N$,
and 2) first  measurements of the two-particle-coincidence spectra, as  a function of:
i) the sum of  kinetic energies $E_n+E_N\equiv E_{nN}$, $S_{N}(E_{nN})$, ii) the opening angle
$\theta_{nN}$, $S_{N}(\cos\theta_{nN})$, and  iii) the center of mass (c.m.)  momentum
$P_{nN}=|\textbf{P}_{nN}|$, $S_{N}(P_{nN})$ with
$\textbf{P}_{nN}=\textbf{p}_n+\textbf{p}_N$. On the theoretical side this  implies
a new challenge for nuclear models which now
have to explain, not only the $1N$- and $2N$-NM decay rates, but also
the shapes and magnitudes of all these spectra, testing in this way both the
 kinematics and the dynamics.

The measured spectra are obtained by counting the numbers of
emitted nucleons $\Delta{\rm N}_N$ within
the energy  bin $\Delta{E}=10$ MeV, or the angular bin  $\Delta\cos\theta=0.05$ always
 corrected by the detection efficiency.  Here we  take advantage of the
fact that  in  the KEK
experiments~\cite{Ki09,Ki06, Bh10} $\Delta{\rm N}_N$ are
 normalized to the number of NM processes ${\rm N}_{NM}$, while for  the
FINUDA proton  spectra \cite{Ag10,Ga10}
we have at our disposal also the number of produced hypernuclei ${\rm N}_{W}$.
Therefore, we can exploit the following relationships
  \be
\frac{\Delta{\rm N}_N}{{\rm N}_{NM}}=\frac{\Delta{\Gamma}_N^{exp}}    {{\Gamma}_{NM}},
\hspace{0.5cm}\frac{\Delta{\rm N}_N}{{\rm N}_{W}}=\frac{\Delta{\Gamma}_N^{exp}  }     {{\Gamma}_{W}},
\label{3}\ee
where  $\Delta\Gamma_N^{exp}$ is the emission rate of protons
within the experimentally
fixed bin.  Thus,
 \be
\frac{{\rm N}_N}{{\rm N}_{NM}}=\frac{{\Gamma}_N^{exp}
}{{\Gamma}_{NM}}, \hspace{0.5cm}\frac{{\rm N}_N}{{\rm
N}_{W}}=\frac{{\Gamma}_N^{exp}  }{{\Gamma}_{W}}, \label{4}\ee
where  ${\rm N}_N=\sum_{i=1}^m\Delta{\rm N}_N$ is  the total
number of NM events decaying to the mode $N$,
 and $\Gamma_N^{exp}=\sum_{i=1}^m\Delta\Gamma_N^{exp}$,
is the corresponding decay rate; m is the number of bins. We note
that the last relation in \rf{4} agrees with Eq. (13) in
Ref.~\cite{Pa07}, and  that $\Gamma_N^{exp}$ accounts for  both
the $1N$-NM and $2N$-NM  decays as well as for  the FSIs.

 The correspondence between theory and data is
\br \Delta\Gamma_N^{th}&\Longleftrightarrow &\Delta\Gamma_N^{exp}
=\left.{\Gamma}_{NM} \frac{\Delta{\rm N}_N}{{\rm
N}_{NM}}\right|_{KEK} ={\Gamma}_{W} \left.\frac{\Delta{\rm
N}_N}{{\rm N}_{W}}\right|_{FINUDA}. \label{5}\er For
$_\Lambda^{12}$C is ${\Gamma}_{NM}=0.95\pm0.04$~\cite{Ki09}, and
as it was  pointed out in Ref.~\cite{Ag09}
 \be
{\Gamma}_{W}(A)=(0.990\pm 0.094)+(0.018\pm0.010)~A,
\label{6}\ee
for all measured hypernuclei in the mass range $A=4-12$. The
theoretical decay widths are $\Delta\Gamma_N^{th}=S_N(E_N)\Delta
E$,\etc. Below we briefly sketch the corresponding expressions for
spectral densities $S_{N}$  within the IPSM.

\section{IPSM for  the $1N$-NM decay}\label{Sec2}
Within the IPSM
\cite{Ba09,Ba02,Kr03,Ba03,Ba07,Ba08,Ba10,Kr10,Kr10a,Go11,Go11a}: i)
 the initial hypernuclear state is taken as a hyperon $\Lambda$ in
single-particle state $j_\Lambda=1s_{1/2}$ weakly coupled to an
$(A-1)$ nuclear core of spin $J_C$, i.e.,
$\ket{J_I}\equiv\ket{(J_Cj_\Lambda)J_I}$, ii)
the nucleon ($N=p,n$) inducing the decay  is in the single-particle state
$j_N$ ($j\equiv nlj$), iii) the final  residual nucleus states are:
$\ket{J_F}\equiv\ket{(J_Cj_N^{-1})J_F}$, iv)
 the liberated energy is
\be \Delta^j_{N} =\Delta  + \varepsilon_{\Lambda} +
\varepsilon^j_{N}, \label{7}\ee
 where  $\varepsilon$'s are
single-particle energies, and $\Delta=\Mass_\Lambda-\Mass$, and v)
the   c.m. momenta, and relative momenta  of the emitted particles
are:
 \br
P_{nN}&=&\sqrt{(A-2)(2\Mass\Delta^j_{N}- p_n^2 -p_N^2)},
~~p_{nN}=\sqrt{\Mass \Delta^j_{N}- \frac{A}{4(A-2)} P_{nN}^2}.
 \label{8}\er
It follows that the $1N$-NM decay rate reads
 \br
\Gamma_{N}&=&\sum_{j}\Gamma_{N}^j;~~~\Gamma_{N}^j=\int
\I^j_{N}(p_{nN},P_{nN})d\Omega_{nN}, \label{9}\er
 where $d\Omega_{nN}$ is the phase space factor, which depends on the spectra that one is interested in,
 and
 \be
\I^j_{N}(p_{nN},P_{nN})=\sum_{J=|j-1/2|}^{J=j+1/2}F^j_{NJ}\sum_L\T_{NJL}^j(p_{nN})
\O_L^2(P_{nN}). \label{10}\ee

The information on  nuclear structure is contained in the
spectroscopic factor \br F^j_{NJ}&=&\hat{J_I}^{-2}\sum_{J_F}
|\Bra{J_I}\left( a_{j_N}^\dag a_{j_\Lambda
}^\dag\right)_{J}\Ket{J_F}|^2,\label{11}\er
 where
$\hat{J}=\sqrt{2J+1}$. The values for $J_I$, and $J_C$ are  taken
from experimental data and for hypernuclei of interest here are
listed in Tables I of Ref. \cite{Kr10}. The resulting factors
$F^j_{NJ}$ are presented in Tables II of the same paper.

The kinematics are enclosed in: i) the phase space factor   $d\Omega_{nN}$, and ii)
the overlap
\be
 \O_L(P_{nN}) =\int R^2dRj_L(P_{nN}R){\rm R}_{0L}(b/\sqrt{2},R),
 \label{12}\ee
 between  the c.m. radial wave functions
${\rm R}_{0L}$  of the bound particle and $j_L$  of the  outgoing particle,
where $b$ is the harmonic oscillator  size parameter.

 The decay dynamics is contained in
\br
\T_{NJL}^j(p_{nN})&=&\sum_{Sl\lambda T}|\M(p_{nN};lL\lambda
SJ\TI;{j_\Lambda j_N J\tLN})|^2, \label{13}\er
where
$\bm{\lambda}=\bm{l}+\bm{L}$, with $\bm{l}$ and $\bm{L}$ being,
respectively, relative and  c.m. angular momenta,
 $\TI\equiv \{TM_T,M_T=m_{t_\Lambda}+m_{t_N}\}$, and $\tLN\equiv \{t_\Lambda=1/2,m_{t_\Lambda}=-1/2,
t_N=1/2,m_{t_N}\}$, with
$m_{t_p}=1/2$,  and $m_{t_n}=-1/2$, where   we have  assumed
that the $\Lambda N\go nN$ interaction occurs with the isospin
change $\Delta T=1/2$. Moreover
\begin{eqnarray}
&&\M(p_{nN};lL\lambda SJ\TI;{j_\Lambda j_N J\tLN})
=\sqi\left[1-(-)^{l+S+T}\right] \nonumber \\ &\times& ({lL\lambda
SJ\T}|V(p_{nN})|{j_\Lambda j_N J\tLN}), \label{14}
\end{eqnarray}
 where (and henceforth) the ket $|)$, unlike  $\ket{}$, indicates that the
state is not antisymmetrized, and  $V(p_{nN})$ is the transition potential.

To find  the kinetic energy $E_{N}$ and opening angle $\theta_{nN}$ spectra it is convenient
to express the  single-particle transition rate as
 \br
\Gamma_{N}&=&(A-2)\frac{8\Mass^3}{\pi}\sum_{j_N} \int _{-1}^{+1}d\cos\theta_{nN}
\int_0^{{\tilde E}_{j_N}} dE_{N}\sqrt{\frac{E_{N}}{E_N'}}\, E_n\, \I_{j_N}(p_{nN},P_{nN}),
 \label{15}\er
with
 \br
E'_N&=&(A-2)(A-1)\Delta_{j_N}-E_{N} [(A-1)^2-\cos^2\theta_{nN}],
\label{16}\er
  \br
  E_n&=&\left[\sqrt{E'_N} -\sqrt{E_{N}}\cos{\theta_{nN}}\right]^2(A-1)^{-2},
\label{17}\er
 and
\be {\tilde E}_{j_N}=\frac{A-1}{A}\Delta_{j_N}. \label{18} \ee
Throughout the integration one has to enforce the condition $E'_N
> E_{N}\cos^2\theta_{nN}$. It might be worth noticing that, while
$E'_N$ does not have a direct physical meaning, $E_n$ is the
energy of the neutron that is the decay-partner of the nucleon $N$
with energy $E_N$. The corresponding spectra read \br
S_N(E_{N})&=&\frac{d\Gamma_N}{dE_N}=
(A-2)\frac{8\Mass^3}{\pi}\sum_{j_N}\int _{-1}^{+1}
d\cos\theta_{nN}\sqrt{\frac{E_{N}}{E_N'}}\, E_n\,
\I_{j_N}(p_{nN},P_{nN}), \label{19}\er
 and
 \br
S_N(\cos\theta_{nN}) &=&\frac{d\Gamma_N}{d\cos\theta_{nN}}=
(A-2)\frac{8\Mass^3}{\pi}\sum_{j_N} \int_0^{{\tilde E}_{j_N}}
dE_{N}\sqrt{\frac{E_{N}}{E_N'}}\, E_n \I_{j_N}(p_{nN},P_{nN}).
\label{20}\er
As  one-proton (one-neutron) induced decay prompts
the emission of an $np$ ($nn$) pair the total neutron kinetic
energy spectrum is: \be
S_{nt}(E_n)=\frac{d\Gamma_{nt}}{dE_n}\equiv S_p(E_n)+2S_n(E_n),
\label{21}\ee where $\Gamma_{nt}=\Gamma_p+2\Gamma_n$ is the total
neutron transition rate.

Similarly, we get
\br
S_N(E_{nN})&=&
\frac{4\Mass^3}{\pi}\sqrt{A(A-2)^3}\sum_{j_N}
\sqrt{(\Delta_{j_N}-E_{nN})(E_{nN}-{\tilde\Delta}_{j_N})}
 \I^j_{N}(p_{nN},P_{nN}),
\label{22}\er for the kinetic energy sum  $E_{nN}=E_n+E_N$, and
\br S_N(P_{nN})&=&
\frac{2\Mass}{\pi}\sqrt{\frac{A}{A-2}}\sum_{j_N}
P_{nN}^2\sqrt{P_{j_N}^2-P_{nN}^2}
\I_{j_N}(p_{nN},P_{nN}),\label{23}\er for the c.m. momentum
$P_{nN}$ spectra, with \be
{\tilde\Delta}_{j_N}=\Delta_{j_N}\frac{A-2}{A},~~\mbox{and}~~
{P}_{j_N}=2\sqrt{\frac{A-2}{A}\Mass\Delta_{j_N}}
 \label{24}\ee
 being, respectively, the  maximum values of $E_{nN}$, and  $P_{nN}$ for each $j_N$.

 The  outline of the numerical calculation is the following:  a) The transition potential for  the emission of the
 $nN$ pair, $V(p_{nN})$   in Eq. \rf{14},
is  described  by two  OME models with the weak coupling constants
from Ref.~\cite{Pa97,Pa02}, namely: M1) only the $\pi+K$ exchange
is considered, and M2) the full  $\pi+K+\eta+\rho+\omega+K^*$ is
taken into account, b) The  initial and final SRCs,  as well as
the finite nucleon size effects are included in the same way as in
our previous
works~\cite{Ba09,Ba02,Kr03,Ba03,Ba07,Ba08,Ba10,Kr10,Kr10a}, and
c) The parameter $b$ in Eq. \rf{12} is evaluated as in
Ref.~\cite{Kr03}, \ie $b=1/\sqrt{\hbar \omega M_{\rm N}}$, with
$\hbar \omega =45A^{-1/3}-25A^{-2/3}$ MeV.

\section{KEK Experiment}
\begin{figure}[htpb]
\begin{center}
\includegraphics[width=0.6\linewidth]{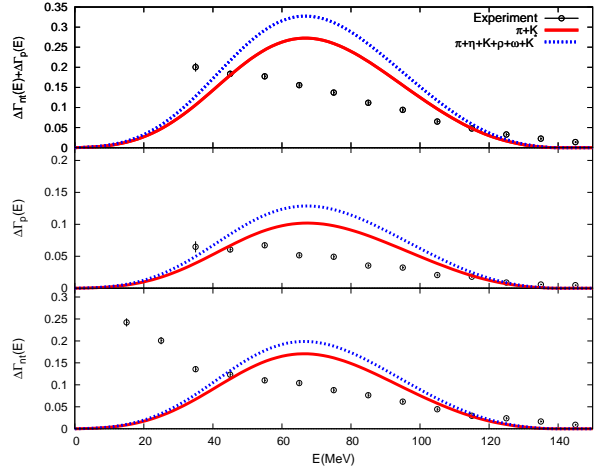}
\vspace{.5cm} \caption{\label{F2}Comparison between the
experimental and theoretical kinetic energy spectra for neutrons
(lower panel), protons (middle panel), and the sum of both (upper
panel). The KEK  experimental data are from \cite{Ki09,Bh10}, and
the relation \rf{5}
 has been used.
 The theoretical results have been evaluated from the relations
$\Delta \Gamma_p(E)=S_p(E)\Delta E$, and
 $\Delta \Gamma_{nt}(E)=[S_p(E)+2S_n(E)]\Delta E$,
where $\Delta E=10$ MeV is the experimental energy bin, and
$S_N(E_{N})$ are given by \rf{19}. Two OME potentials have been
used:
 M1) only $\pi+K$ exchange is considered, and M2) full $\pi+K+\eta+\rho+\omega+K^*$ exchange
 is taken into account.}
\end{center}
\end{figure}

\begin{figure}[htpb]
\vspace{2.cm}
\begin{center}
\includegraphics[width=0.6\linewidth]{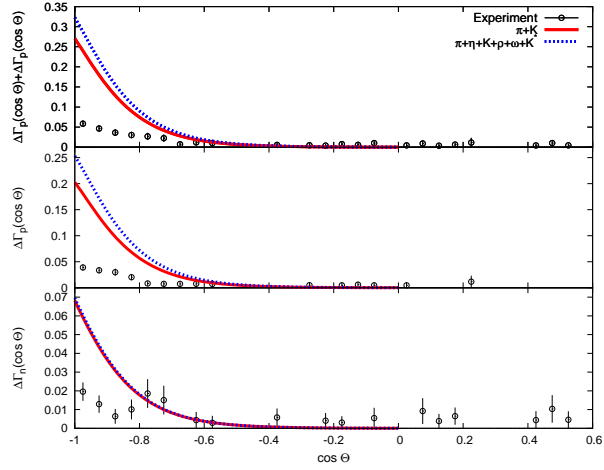}
\vspace{.5cm} \caption{\label{F3}Comparison between experimental
and calculated opening angle correlations for neutron-neutron
pairs (lower panel),  proton-neutron pairs (middle panel),
 and the sum of both (upper panel).
 The KEK experimental data are from \cite{Ki09,Bh10}, and the relations \rf{5}
 have been used.
 The theoretical results have been evaluated from the relations
$\Delta \Gamma_N(\cos\theta_{nN})=S_N(\cos\theta_{nN})\Delta
\cos\theta$, where $\Delta \cos\theta=0.05$  is the experimental
opening angle bin, and $S_N(\cos\theta_{nN})$ are given by
\rf{20}. The same OME potentials have been used as in Fig.
\ref{F2}.}
\end{center}
\end{figure}
\begin{figure}[htpb]
\begin{center}
\includegraphics[width=0.6\linewidth]{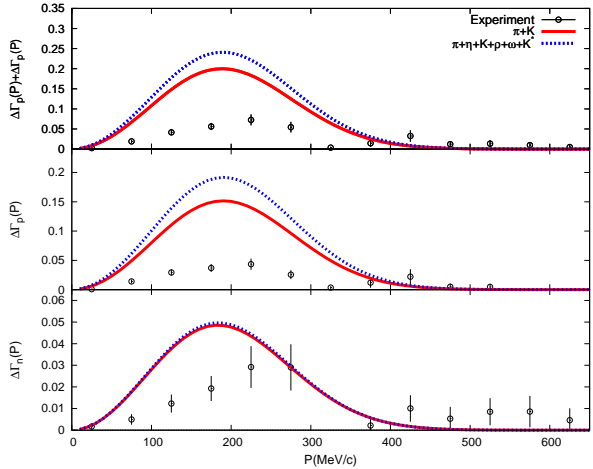}
\vspace{0.5cm} \caption{\label{F4}Comparison between experimental
and calculated c.m. momentum for neutron-neutron pairs (lower
panel),  proton-neutron pairs (middle panel),
 and the sum of both (upper panel).
 The KEK experimental data are from \cite{Ki09,Bh10}, and  relations \rf{5}
 have been used.
 The theoretical results have been evaluated from the relations
$\Delta \Gamma_N(P_{nN})=S_N(P_{nN})\Delta P$, where $\Delta P=50$
MeV$/$c is the experimental center of mass momentum bin, and
$S_N(P_{nN})$ are given by \rf{23}. The same OME potentials  have
been used as in Fig. \ref{F2}.}
\end{center}
\end{figure}

In Figs. \ref{F2}, \ref{F3}, and \ref{F4}   we compare the
KEK~\cite{Ki09,Bh10} experimental data with  the IPSM
calculations. As seen from  Fig.  \ref{F2} the calculated proton
and neutron kinetic energy spectra
 agree  fairly well  with data for $E_N>40$ MeV, and specially
when  the $\pi+K$ potential is used, in  spite of not considering
the FSIs.
 This is not the case, however,  for the angular distributions, and the c.m. momenta
of the $nN$ pairs shown in Figs. \ref{F3}, and \ref{F4}, where the
theory overestimates the data quite significantly, and
particularly for the $np$ pairs.
 It could be worth noticing that both potentials yield very similar results
 for the $nn$ pairs.

The  experimental  decay rates $\Gamma_N$ are obtained  by summing
over all  $\Delta\Gamma_N$ in Figs. \ref{F2}, \ref{F3}, and
\ref{F4}. The results are shown in the first three rows of Table
\ref{tab1}.
 In the following  two rows
the results of the IPSM  calculation are exhibited. No energy or angular cutoffs
are done on the calculated results, since their effects are insignificant. It is self
evident that the theory overestimates the data and in particular so
in the case M1. We are aware that a more elaborate description  of the SRCs,
as  performed in Ref.~\cite{Pa02},
could improve the agreement between data and theory for transition rates which are shown in
Table \ref{tab1}. In this regard the FSIs
could  also help, which in
addition are able to modify the spectra by  redistributing  the calculated  transition
strengths.
On the other hand  the new $2N$-NM channel can only increase the theoretical results
but never decrease them, and thus the discrepancy with experiment
 would be still more pronounced.

\begin{table}[htpb]
\caption{Comparison between the experimental  transition rates
derived from Figs. \ref{F2}, \ref{F3}, and  \ref{F4}, and the IPSM
calculation, making use of two  OME potentials: M1) exchange of
$\pi+K$ only, and M2)  full $\pi+K+\eta+\rho+\omega+K^*$. The
results for Fig. \ref{F2} correspond to  the proton and neutron
thresholds $E_N\ge 35$ MeV. } \label{tab1}
\bigskip
\begin{center}
\begin{tabular}{|c|ccccc|}
\hline
Source &$\Gamma_{p}$&$\Gamma_{n}$&$\Gamma_{p}+\Gamma_{n}$&$\Gamma_{nt}$&$\Gamma_{p}+\Gamma_{nt}$\\
\hline
Fig. \ref{F2} &$0.419\pm0.013 $&$-$&$-$&$0.823\pm 0.016  $&$1.242 \pm0.021$\\
Fig. \ref{F3} &$0.197 \pm0.024$&$0.147\pm0.022$&$0.345\pm0.033$&$-$&$- $\\
Fig. \ref{F4} &$0.197\pm0.023$&$0.136\pm0.022$&$0.338\pm0.033$&$-$&$- $\\
M1             &$0.627$&$0.201$&$0.828$&$1.455$&$ 1.656$\\
M2       &$0.792 $&$0.205$&$0.997$&$1.789$&$ 1.994 $\\
  \hline
   \end{tabular}
\end{center}
\end{table}

The measurement  of $\Gamma_{2}$ in Ref.~\cite{Ki09}
 is based on  INC-$1N$ calculations that overestimate the data above thresholds $E_N>30$ MeV
for both single and coincidence spectra. The authors  denominate this difference
as "quenching"  of  data in relation to theory~\cite{Ki09,Bh07}. Moreover,  they
attribute it to the lack of the $2N$-NM channel, which when
incorporated, in proportion of $\Gamma_{2}/\Gamma_{NM}=0.29\pm 0.13$,
 decreases the calculated  decay rate $\Gamma_{NM}$
and yields  agreement with data. Several observations are pertinent here:

1) Usually the word "quenching" has a   different meaning in nuclear physics.
 The emblematic example is that of  Gamow-Teller (GT) strength, the experimental value
 of which,
  either from nuclear $\beta$-decay or from charge-exchange reactions, is
  in general overestimated by  calculations. This discrepancy is often  resolved  by
 including additional degrees of freedom;
 neither the total theoretical value  of the GT operator $\sigma\tau$ is modified
nor  opening of new decay or reaction channels is required.

2)
The present calculation also overestimates the data.
Nevertheless, we do not dare to state  that the NM
transition  operator $V(p_{nN})$ is quenched,  since as pointed out previously
its evaluation involves many uncertainties,
 which is not the case with the GT operator.

3) The  INC sequence in the calculations of the FSIs
performed so far~\cite{Go11,Go11a,Ga03,Bau10} is always triggered
by the primary $nN$ or $nNN$ nucleons. These nucleons  are produced by the
same OME dynamics as we have used here.
In the KEK calculation the primary NM weak dynamics is totally ignored~\cite{Ki09}.

4) The Monte Carlo study of $_\Lambda^{12}$C performed in
Refs.~\cite{Go11,Go11a}, where the production of  primary nucleons is described
according to IPSM, does not show any evidence of the "quenching"
reported  in Ref.~\cite{Ki09}. This fact suggests that the
uncertainties in the FSIs calculation must be carefully considered
before extracting definite conclusions from  comparison of
the INC results with  the experimental data.

5) The above mentioned uncertainties embedded in the KEK description of the primary NM decay
dynamics cannot but be augmented when the FSIs are considered. Therefore, the ambiguities  involved in INC-$1N$ calculation are probably as large or still larger
than the observed "quenching".

Briefly, in our opinion the reckoning followed by the  KEK group is not sufficiently robust to support the
statement that the NM $nNN$-decay mode  has been unequivocally observed experimentally.

\section{FINUDA Experiment}

\begin{figure}[htpb]
\hspace{-0.5cm}
\begin{tabular}{cc}
\includegraphics[width=7cm,height=6.cm]{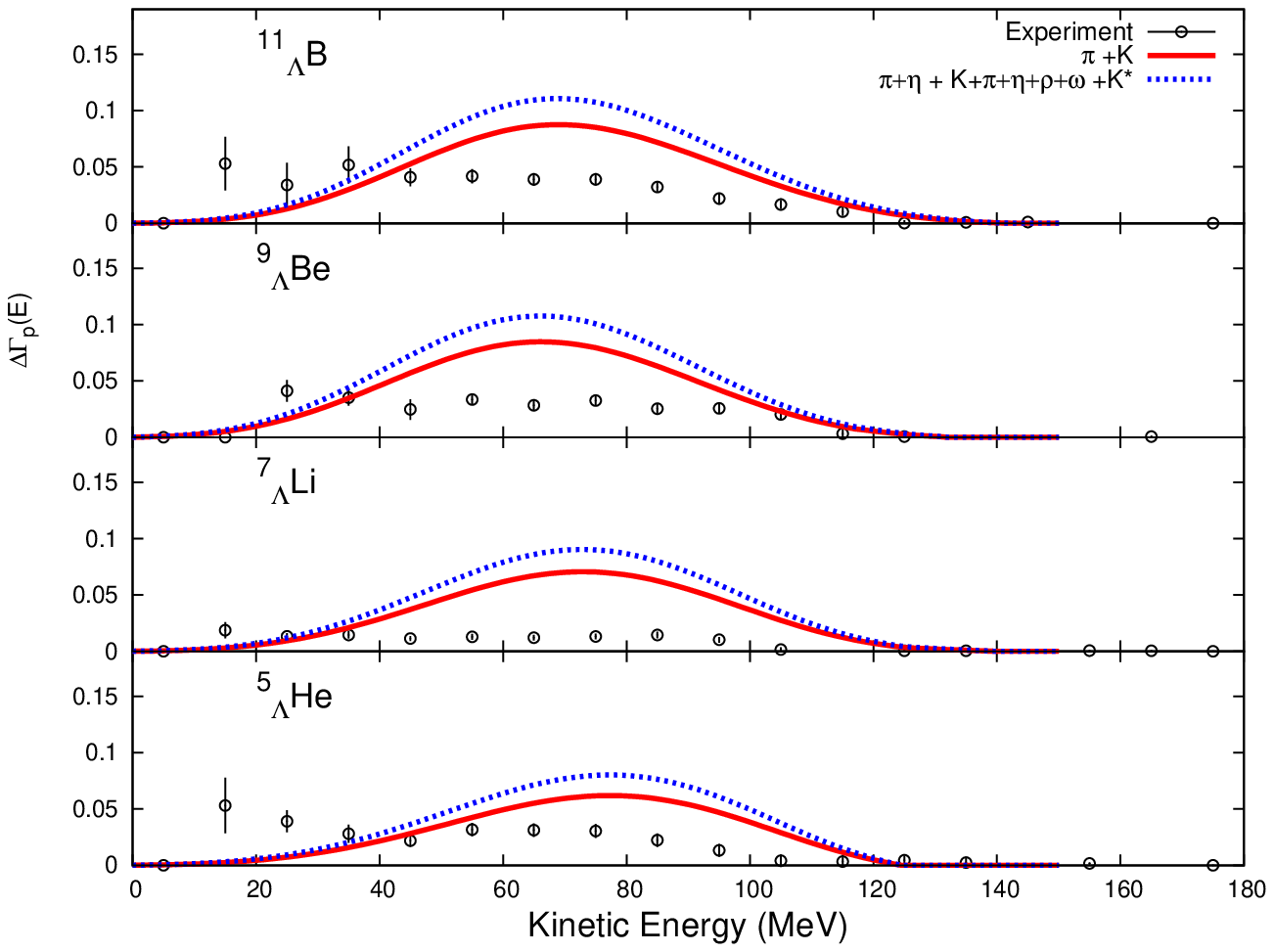}&
\includegraphics[width=7cm,height=6.cm]{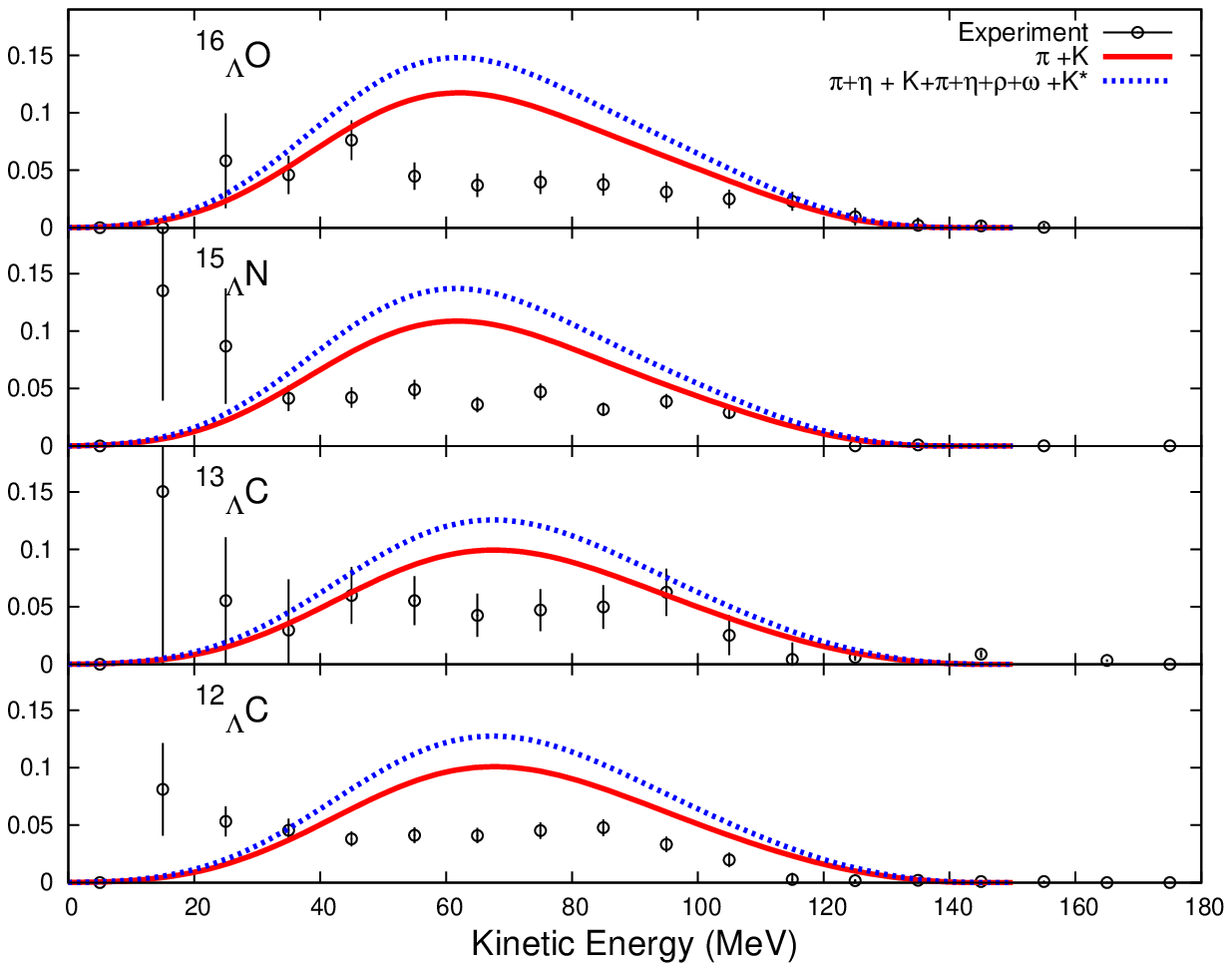}
\end{tabular}
\vspace{0.51cm} \caption{\label{F5} Experimental
data~\cite{Ag10,Ga10} for proton kinetic energy spectra are
compared with the  IPSM results.
 Theoretical results have been evaluated from the relation
$\Delta \Gamma_p(E)=S_p(E)\Delta E$, where $\Delta E=10$ MeV is
the experimental energy bin. The same OME potentials have been
used as in Fig. \ref{F2}.}
\end{figure}

The comparison between
 FINUDA~\cite{Ag10,Ga10} experimental data
and the IPSM  calculation for the proton kinetic energy spectra is
shown in Fig. \ref{F5}. One sees that the theory overestimates the
data significantly,  particularly in the case of $_\Lambda^{7}$Li.
Similarly as in the case of the KEK data,
 discrepancies between data and theory, are manifested also
in the case of FINUDA transition rates $\Gamma_{p}$, as can be seen from Table
\ref{tab2}.

\begin{table}[htpb]
\caption{In the first three columns the comparison is done between the experimental
 transition rates $\Gamma_{p}$ derived from Figs. \ref{F5} and the
 IPSM  calculation, making use of two  OMEP:
 M1)  full $\pi+K+\eta+\rho+\omega+K^*$, and M2)  exchange of  $\pi+K$ only.
 In the last two columns are listed the  mean values of the Gaussian fit as obtained in
Ref.~\cite{Ag10}, and calculated maxima within the IPSM, in units of MeV.} \label{tab2}
\bigskip
\begin{center}
\begin{tabular}{|c|ccc|cc|}
\hline
Hypernucleus    &$\Gamma_{p}^{\mbox{\tiny FINUDA}}$&$\Gamma_{p}^{\mbox{\tiny M1}}$
&$\Gamma_{p}^{\mbox{\tiny M2}}$&${\hat E}^{\mbox{\tiny FINUDA}}$&${\hat E}^{\mbox{\tiny IPSM}}$\\
\hline
$_\Lambda^{5}$He &$0.284$&$0.466$&$0.360$&$68.5\pm4.1$&$76.7$\\
$_\Lambda^{7}$Li &$0.123$&$0.531$&$0.415$&$76.7\pm5.2$&$72.3$\\
$_\Lambda^{9}$Be &$0.269$&$0.627$&$0.494$&$78.2\pm6.2$&$66.2$\\
$_\Lambda^{11}$B &$0.380$&$0.667$&$0.527$&$75.1\pm5.0$&$68.9$\\
$_\Lambda^{12}$C &$0.358$&$0.792$&$0.627$&$80.2\pm2.1$&$67.5$\\
$_\Lambda^{13}$C &$0.437$&$0.776$&$0.614$&$83.9\pm12.8$&$67.6$\\
$_\Lambda^{15}$N &$0.402$&$0.821$&$0.651$&$88.1\pm6.2$&$61.6$\\
$_\Lambda^{16}$O &$0.320$&$0.906$&$0.718$&$93.1\pm6.2$&$62.0$\\
  \hline
 \end{tabular}
\end{center}
\end{table}

\begin{table}[h]
\centering \caption{Results for  the $\chi^{2}$ parameters $a$ and
$b$, and the corresponding ratios ${\Gamma_{2}}/{\Gamma^0_{p}}$, and
$\Gamma_{2}/\Gamma_{\rm NM}$;  A) Ref.~\cite{Ag10}, B) data from Ref.~\cite{Ag10},
and the peak energies ${\hat E}^{\mbox{\tiny IPSM}}$ from last column in Table \ref{tab2}
instead of ${\hat E}^{\mbox{\tiny FINUDA}}$, and C) same as in
B but performing the fit with Eq. \rf{30} instead of Eq. \rf{26}.} \label{tab3}
\bigskip
\begin{tabular}{|c|c|c|c|c|}
\hline case&$a$&$b$&${\Gamma_{2}}/{\Gamma^0_{p}}$&$\Gamma_{2}/\Gamma_{\rm NM}$ \\
\hline
A&$0.654\pm0.138$&$  0.009\pm  0.013 $&$ 0.43\pm 0.25$&$ 0.24\pm 0.10$\\
B&$  0.674\pm  0.108$&$ -0.008\pm  0.009$&$  0.53\pm  0.51$&$  0.26\pm  0.10$\\
C&$  0.135\pm  0.408$&$  0.207\pm  0.202$&$ -0.42\pm  0.14$&$ -0.40\pm  0.10$\\
  \hline
   \end{tabular}
\end{table}

The determination of the  branching ratio $\Gamma_{2}/\Gamma_{NM}$ in Ref.~\cite{Ag10} is based
on: i)  separating the total number of detected protons ${\rm N}_p$ into  low, and high energy
regions populated by  ${\rm N}_p^<$, and  ${\rm N}_p^>$ protons, respectively,
 and ii) assuming
that the $1N$-NM decay rates of the primary protons ${\rm N}^0_p\equiv {\rm N}(\Lambda p\go np)$
 are equal in these two regions, \ie
$\Gamma_p^{0,<}=\Gamma_p^{0,>}=\Gamma^0_p/2$, where the superscript
 $0$ indicates that this decay rate, unlike those in Tables \ref{tab1} and \ref{tab2},
doesn't contain contributions coming from the FSIs and $2N$-NM decay.
 Moreover, they write
\br
R&\equiv&\frac{{\rm N}_<}{{\rm N}}
=\frac{0.5+\Gamma_2/\Gamma^0_p+N_<^{FSI}/{\rm N}^0_p}
{1+\Gamma_2/\Gamma^0_p+N^{FSI}/{\rm N}^0_p},
\label{25}\er
where ${\rm N}^{FSI}={\rm N}_<^{FSI}+{\rm N}_>^{FSI}$.
The partition energies ${\hat E}$ are fixed  as being  the mean   values of   Gaussian-function fits
of  each proton spectrum from $80$ MeV onwards, and  the experimental values of the ratios
$R\equiv{\rm N}_p^</{\rm N}_p$ found in this way are approximated
by a linear function of the mass number  $A$, \ie
\be
R(A)=a+bA,~~\mbox{where}~~a=\frac{0.5+\Gamma_2/\Gamma^0_p}
{1+\Gamma_2/\Gamma^0_p},
\label{26}\ee
 does not depend on $A$, and yields
\be
 \frac{\Gamma_{2}}{\Gamma^0_{p}}=\frac{a-0.5}{1-a},~~
  \frac{\Gamma_{2}}{\Gamma_{\rm NM}}
  =\frac{a-0.5}{(1-a)\Gamma^0_n/\Gamma^0_p+0.5}.
\label{27}
\end{equation}
Finally,   a $\chi^{2}$ fit  for $R(A)$ was done obtaining the values of   $a$, and  $b$
shown in the  row A of Table \ref{tab3}, together with the
resulting  $\Gamma_{2}/\Gamma^0_{p}$, and $\Gamma_{2}/\Gamma_{NM}$
derived from \rf{27} with  $\Gamma^0_n/\Gamma^0_p=0.48 \pm 0.08$, which is
the mean value  of recent experimental results~\cite{Bh07}.

The manner of deducing  the $2N$-NM decay rate as performed in Ref.
~\cite{Ag10}  can be challenged
from several aspects which we  point out below.

\begin{figure}[htbp]
\begin{center}
\includegraphics[width=65mm]{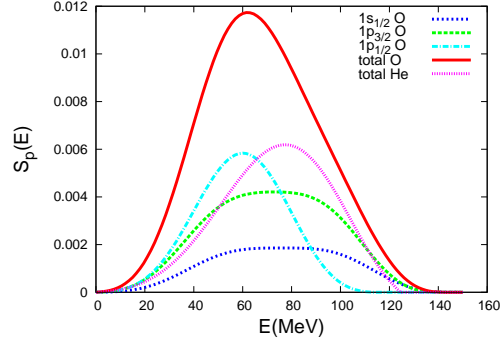}
\caption{\label{F6} Calculated $^{5}$He and  $^{16}$O spectra
within the IPSM employing the $\pi+K$ transition potential.
Contributions of different partial waves $s_{1/2}$, $p_{3/2}$, and
$p_{1/2}$
 to the total $^{16}$O spectra are also displayed.}
\end{center}
\end{figure}

1) How to arrive  from \rf{25} to \rf{26} is not a trivial issue. One possibility is to neglect
the last term in the denominator of \rf{25}, arguing, as was done in Ref. \cite{Ag10}, that the
FSIs tend to remove protons from the high energy part of the spectrum (${\rm N}_>^{FSI}<0$)
while filling the low energy region (${\rm N}_<^{FSI}>0$), with the net result that
${\rm N}^{FSI}={\rm N}_<^{FSI}+{\rm N}_>^{FSI}\cong 0$. Therefore
\begin{figure}[htpb]
\hspace{-0.5cm}
\begin{tabular}{cc}
\includegraphics[width=7cm,height=6.cm]{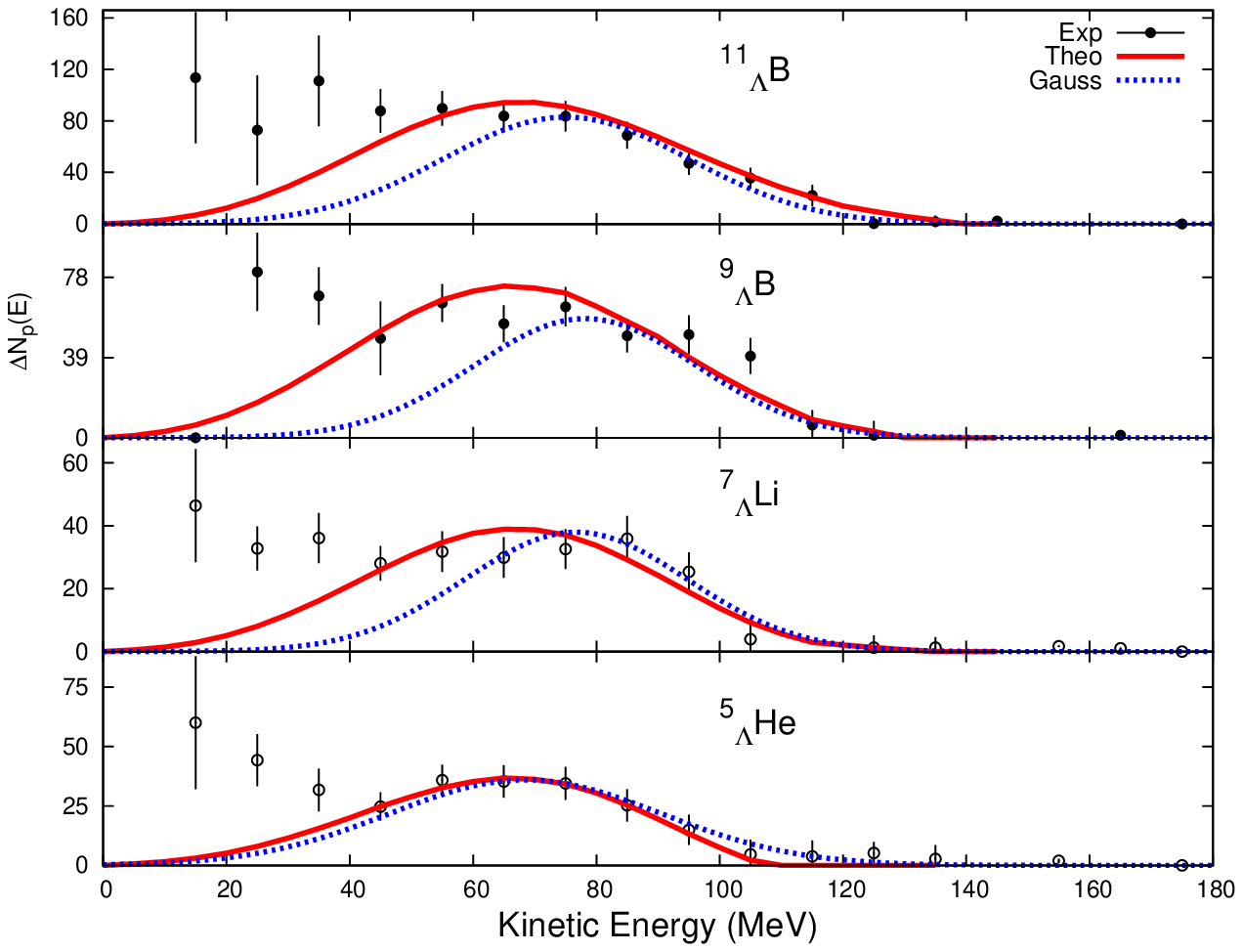}&
\includegraphics[width=7cm,height=6.cm]{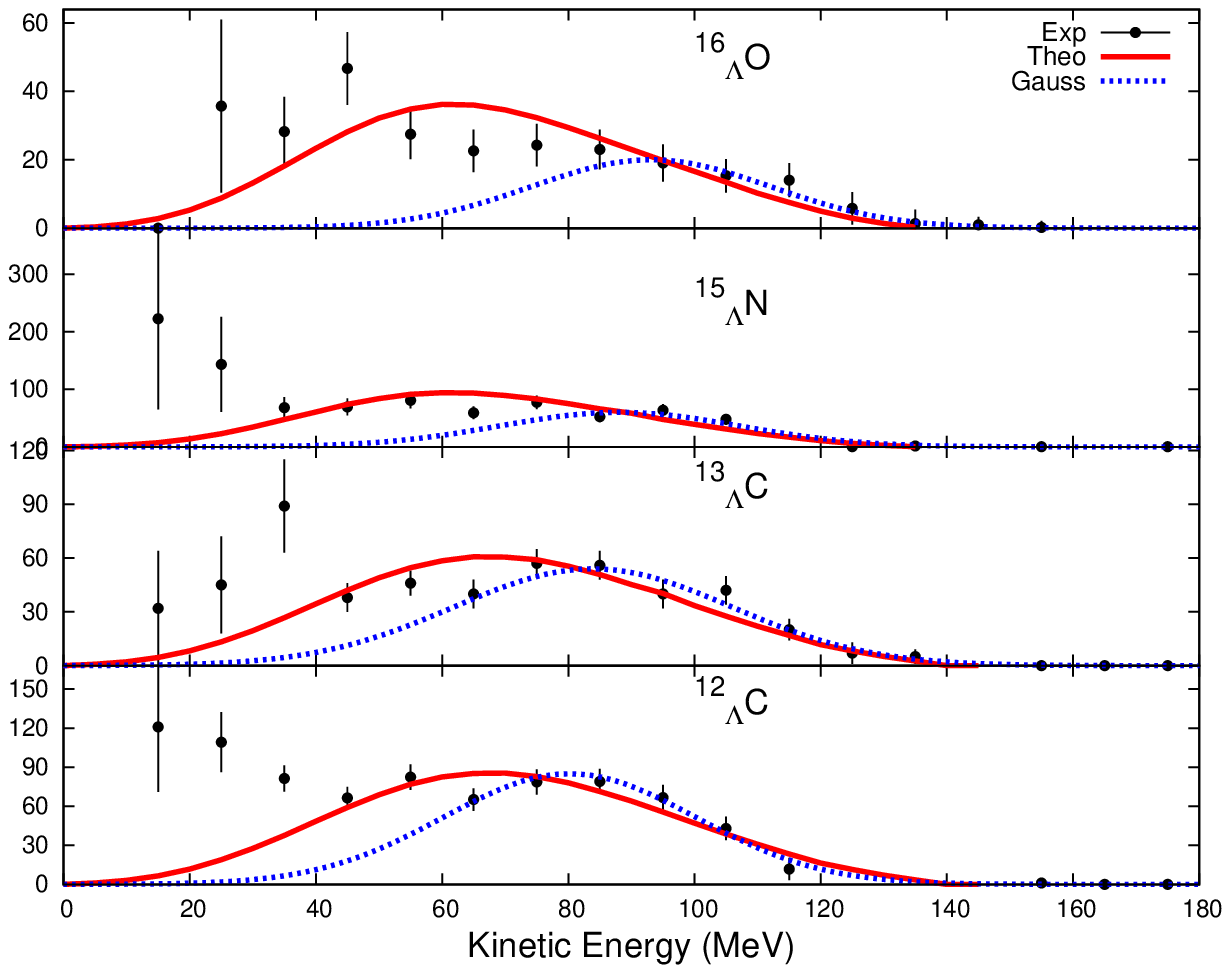}
\end{tabular}
\vspace{0.3cm} \caption{\label{F7} Experimental data and Gaussian
fits from Ref~\cite{Ag10} for proton kinetic
 energy spectra, are compared with the  IPSM results  normalized for proton energy $E_p\ge 40$ MeV. }
\end{figure}
\br
R&\cong&a+\frac{{\rm N}_<^{FSI}}{(1+\Gamma_2/\Gamma^0_p){\rm N}^0_p}
\cong a+\frac{{\rm N}_<^{FSI}}{{\rm N}^0_p},
\label{28}\er
which, when compared with \rf{26}, yields
\be
{{\rm N}_<^{FSI}}\cong b{A}{\rm N}^0_p.
\label{29}\ee
At first glance this sounds
reasonable because the effect of FSIs should increase with $A$.
But, since  $b=0.009\pm 0.013$ one finds that the number of particles  ${\rm N}_<^{FSI}$ per nucleon
that are dislocated from high to low energy  is less than $1\%$ of the total number
of primary protons produced by the $1N$-NM decay,
 which is unrealistic.

2) The assumption $\Gamma_p^{0<}=\Gamma_p^{0>}=\Gamma^0_p/2$ is badly supported by the
 IPSM,  where  $\Gamma_p^{0>}$ becomes  progressively larger than  $\Gamma_p^{0<}$
  as the mass number is increased.
Why it is so can be understood from  inspection of Fig. \ref{F6},
where the spectra of $_\Lambda^{5}$He and $_\Lambda^{16}$O  are
displayed. In the case of
 $_\Lambda^{16}$O the
partial wave contributions are also presented.
One sees that, while in $_\Lambda^{5}$He
only the  orbital $s_{1/2}$ contributes, being
$\Gamma_p^{0<}$ appreciably  larger than $\Gamma_p^{0>}$
 ($\Gamma_p^{0<}=0.55\Gamma^0_p$), for other hypernuclei  also the orbitals
$p_{3/2}$, and $p_{1/2}$ contribute. Their  spectra are wider and localized at higher energies
than the corresponding $s_{1/2}$ strength.
This causes  the total
transition strength to be gradually shifted  towards higher energies as $A$ increases,
becoming $\Gamma_p^{0<}=0.43\Gamma^0_p$ for $_\Lambda^{16}$O.

\begin{figure}[h]
\begin{tabular}{cc}
\includegraphics[width=0.40\linewidth,clip=]{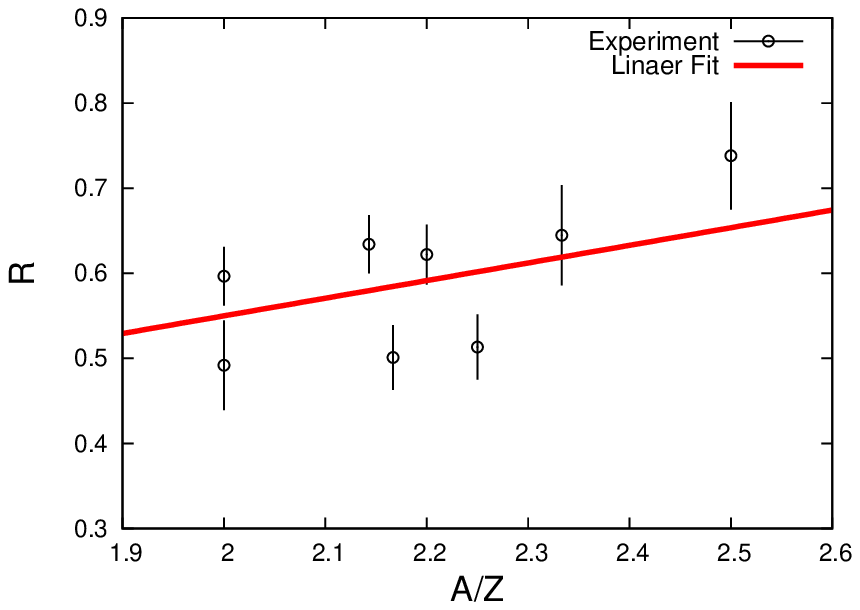}&
\includegraphics[width=0.40\linewidth,clip=]{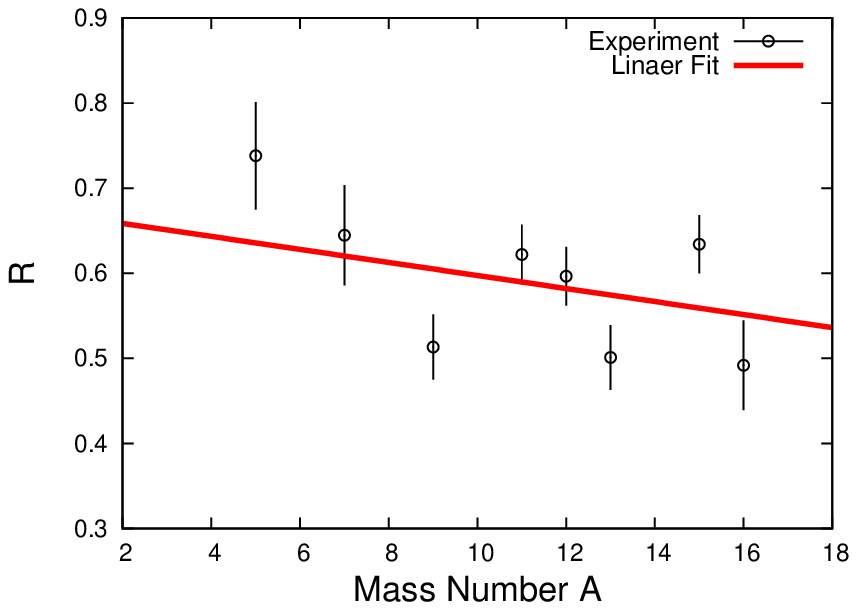}
\end{tabular}
\caption{\label{F8} Ratio $R$ as a function of  $A$ (left panel),
and of $A/Z$ (right panel). In both cases the energies ${\hat
E}^{\mbox{\tiny IPSM}}$, and the corresponding $R\equiv{\rm
N}_</{\rm N}$  are used.}
\end{figure}
3)  Except for $_\Lambda^{5}$He, the calculated proton spectra
 deviate significantly from the Gaussian fits adopted in Ref.~\cite{Ag10}
 for the $1N$-NM decay. Although bell shaped they differ both in energies ${\hat E}$
and in  FWHM widths. This can be observed from  Table \ref{tab2}, and
 Fig. \ref{F7},
where the  IPSM results are  normalized to the experimental data for proton energies $E_p\ge 40$
MeV.  Both OMEP models yield the same
  normalized spectra, since the spectrum shapes depend only on  kinematics.

4) Large differences between  ${\hat E}^{\mbox{\tiny FINUDA}}$, and ${\hat E}^{\mbox{\tiny IPSM}}$,
shown in Table \ref{tab2}, give rise to large differences between values of  ratio
$R$ in  Ref.~\cite[Fig. 2]{Ag10}, and our values  exhibited in
 Fig. \ref{F8}. On the left and right panels of this figure are shown
 the  linear  $\chi^2$ fits of $R$ given by  \rf{26}, and by
\br
R(A,Z)&=&a+bA/Z,
\label{30}\er
respectively.
The later is also a possible parametrization for \rf{25}, as  suggested by  expression \rf{28}
 since  ${\rm N}^0_p$
should be proportional to the number of protons $Z$~\cite{Kr10a}.
The resulting values of   $a$, $b$, ${\Gamma_{2}}/{\Gamma^0_{p}}$,
and $\Gamma_{2}/\Gamma_{\rm NM}$
are listed  listed in rows B and C of Table \ref{tab3}. One sees  that
$a$,  and therefore  ${\Gamma_{2}}/{\Gamma^0_{p}}$, and $\Gamma_{2}/\Gamma_{\rm NM}$
 are very similar in cases A and B.  However, the value of parameter $b$ is
 negative in case B, which makes its physical interpretation
  still more embarrassing than in  case A, since  from \rf{29} it follows that
  the number ${\rm N}_<^{FSI}$
becomes negative, and therefore the FSIs now move particles from  low to high energy.
  Finally, in case C the ratios
${\Gamma_{2}}/{\Gamma^0_{p}}$, and $\Gamma_{2}/\Gamma_{\rm NM}$ turn out to be large and negative,
which obviously doesn't make sense.

We conclude therefore, that similarly to the KEK experiment~\cite{Ki09}, the argumentation
followed by the FINUDA group~\cite{Ag10} can yield several quite different
 results for the ratio $\Gamma_{2}/\Gamma_{NM}$.

\section{Summary and Conclusions}
We have discussed the new  KEK~\cite{Ki09}
 and the FINUDA~\cite{Ag10} experiments on the nonmesonic weak decay
of hypernuclei in the framework of the IPSM, with the dynamics  described by the OPE mechanism,
and using two different models:
 M1) with  $\pi+K$ mesons only, and M2) that comprises  the exchange of the complete pseudoscalar and vector meson octets
($\pi,\eta,K,\rho,\omega,K^*$).
We prefer the first one since the coupling constants $NN\pi$, and $N\Lambda\pi$ are
well established, and the only parameters that are still  uncertain in this case are
the coupling strengths $NNK$,  and $N\Lambda K$. Of course, except for reasons of
 simplicity, there is no justification for not including  the remaining mesons.

 We found that the theory overestimates the data to a great extent. There could be several reasons
 for this, such as: i) the absence of the FSIs in the employed model, ii) the uncertainties
 inherent in the OME model,  iii) the approximation used for the SRCs, \etc.

We argue  that  at present there exist
many uncertainties,  on both  sides experimental and theoretical, which prevent us to draw
definite conclusions regarding the $2N$-NM decay rate $\Gamma_2$.
In view of this situation we are  developing
the IPSM formalism
for this decay channel~\cite{Ba11} that will be  confronted  with the FGM, which is the only
approach used so far~\cite{Ba09}.
Once  this is accomplished it will also enable us to study the effects of
 FSIs on the three-nucleon emission.
New developments in this field of experimentation will be very welcome.
Needless to say that triple ($p, n, n$), and  ($p, p, n$) coincidence measurements
will be extremely useful for   direct measurement of $\Gamma_2$.

After completing  the present work we have learnt that a new derivation of the ratio
$\Gamma_{2}/\Gamma_{\rm NM}$ has been done at FINUDA, based on the
analysis of the ($\pi^-, p, n$) triple coincidence events~\cite{Ag11}. The result is similar
to the previous one~\cite{Ag10}.

\begin{center}
{\bf ACKNOWLEDGEMENTS}
\end{center}
FK is supported by the Argentinean agency CONICET under
contract PIP 0377.  AD is supported by the Brazilian agencies FAPESP and CNPQ.
We are grateful to Eduardo Bauer, Joe Parker, and Gianni Garbarino for
helpful discussion and critical reading of the manuscript.
We also  wish to express our sincere thanks to Gordana Tadi\'{c} for the
meticulous review of the manuscript.

\end{document}